\ifx\pdfoutput\undefined
	\documentclass[dvips,apjl]{emulateapj}
  \DeclareGraphicsExtensions{.eps}
\else
	\documentclass[pdftex,apjl]{emulateapj}
  \DeclareGraphicsExtensions{.pdf}
\fi

\usepackage{amsmath}
\usepackage{apjfonts}

\begin{document}

\title{Exact versus approximate equitemporal surfaces in Gamma-Ray Burst afterglows}

\author{Carlo Luciano Bianco\altaffilmark{1} and Remo Ruffini\altaffilmark{2}}
\affil{ICRA --- International Center for Relativistic Astrophysics.}
\affil{Dipartimento di Fisica, Universit\`a di Roma ``La Sapienza'', Piazzale Aldo Moro 5, I-00185 Roma, Italy.}

\altaffiltext{1}{E-mail: bianco@icra.it}
\altaffiltext{2}{E-mail: ruffini@icra.it}

\begin{abstract}
By integrating the relativistic hydrodynamic equations introduced by Taub we have determined the exact EQuiTemporal Surfaces (EQTSs) for the Gamma-Ray Burst (GRB) afterglows. These surfaces are compared and contrasted to the ones obtained, using approximate methods, by \citet{pm98c,s98,gps99}.
\end{abstract}


\section{Introduction}

The recent explanation of the observed luminosity in X- and $\gamma$-ray energy bands in Gamma-Ray Bursts (GRBs), as well as the comprehension of their spectral properties, depends in an essential way on the determination of the EQuiTemporal Surfaces (EQTSs) in the afterglow era \citep{rbcfx02_letter,Spectr1}. Here we compare and contrasts the exact determination of the EQTSs with the approximate expressions presented by \citet{pm98c,s98,gps99}.

A great deal of consensus has been reached concerning three basic issues in the description of GRB afterglows: \\
{\bf a)} Their origin is generally traced back to the interaction of an ultrarelativistic baryonic matter pulse with the InterStellar Medium (ISM). It is also agreed that the relativistic hydrodynamic equations introduced by Abe \citet{Taub} are the correct theoretical framework to describe such an interaction.\\
{\bf b)} The general definition of the EQTSs is also agreed upon by everyone.\\
{\bf c)} The necessity of determining the boundary conditions of the baryonic matter pulse in the early phases of the afterglow era is also generally recognized.

We illustrate in the following sections the equations needed for the description of these three basic issues and also point out a major difference between our approach and the ones in the current literature concerning the solutions of the Taub equations. We identify in this difference the reason for the results on the EQTSs presented in Fig.~\ref{eqts_comp_ad} and Fig.~\ref{eqts_comp_rad}. Our treatment assumes the exact integration of the Taub equations.

We recall that all the GRB observable quantities depend essentially on the EQTSs. Due care is therefore needed in their correct determination.

\section{The theoretical background}

\subsection{The Taub relativistic hydrodynamic equations}

The relativistic hydrodynamic equations have been expressed by \citet{Taub} in the analysis of the relativistic Rankine-Hugoniot equations, see also \citet{ll} for the explicit comparison between the relativistic and nonrelativistic regimes. Although consensus has been reached on the equations to be used in the description of the GRB phenomenon, differences still exists on the validity of the approximations adopted in the solutions and their physical interpretation. The Taub equations have been used to describe the adiabatic optically thick expansion phase of the pulse generating the GRB \citep[see e.g.][]{mlr93,bkm95,rswx99,rswx00}. In the present context of the GRB afterglow consensus exists that these equations acquire the form \citep[see e.g.][]{bm76,p99,Brasile}:
\begin{subequations}\label{Taub_Eq}
\begin{eqnarray}
dE_{\mathrm{int}} &=& \left(\gamma - 1\right) dM_{\mathrm{ism}} c^2 \label{Eint}\, ,\\
d\gamma &=& - \textstyle\frac{{\gamma}^2 - 1}{M} dM_{\mathrm{ism}}\, , \label{gammadecel}\\
dM &=& \textstyle\frac{1-\varepsilon}{c^2}dE_{\mathrm{int}}+dM_\mathrm{ism}\, ,\label{dm}\\
dM_\mathrm{ism} &=& 4\pi m_p n_\mathrm{ism} r^2 dr \, , \label{dmism}
\end{eqnarray}
\end{subequations}
where $E_{\mathrm{int}}$, $\gamma$ and $M$ are respectively the internal energy, the Lorentz factor and the mass-energy of the expanding pulse, $n_\mathrm{ism}$ is the ISM number density which is assumed to be constant, $m_p$ is the proton mass, $\varepsilon$ is the emitted fraction of the energy developed in the collision with the ISM and $M_\mathrm{ism}$ is the amount of ISM mass swept up within the radius $r$: $M_\mathrm{ism}=(4/3)\pi(r^3-{r_\circ}^3)m_pn_\mathrm{ism}$, where $r_\circ$ is the starting radius of the shock front. In general, an additional condition is needed in order to determine $\varepsilon$ as a function of the radial coordinate. In the following, $\varepsilon$ is assumed to be constant and such an approximation appears to be correct in the GRB context.

\citet{bm76} obtained a self-similar solution of the Taub equations in a so-called ``ultrarelativistic'' approximation with:
\begin{equation}
\gamma_\circ \gg \gamma \gg 1\, ,
\label{g0g1}
\end{equation}
where $\gamma_\circ$ is the initial value of the Lorentz gamma factor of the shock front. In the current literature, such an approximation has been used in order to obtain a simple power-law relation between the Lorentz gamma factor and the radial coordinate of the expanding shock:
\begin{equation}
\gamma \propto r^{-\alpha}\, ,
\label{gamma_app}
\end{equation}
where $\alpha = 3$ corresponds to the fully radiative case (i.e. $\varepsilon=1$ in Eqs.(\ref{Taub_Eq})) and $\alpha = 3/2$ to the fully adiabatic case (i.e. $\varepsilon=0$ in Eqs.(\ref{Taub_Eq})).

In our approach we have exactly integrated the Taub equations. For all the sources which we have analyzed, including the case of GRB 991216 which will be used as a prototype in this paper, we have always found $\gamma_\circ \sim 200$ -- $300$. Under these conditions, we have found that no power-law solutions like Eq.(\ref{gamma_app}) hold in any finite region of the afterglow: at most we can define at every point an ``effective'' power-law index $\alpha_\mathrm{eff}$ which in the early phases of the afterglow is zero, then reaches a maximum value and then decreases all the way back down to zero in the latest phases of the afterglow. Such a maximum value of $\alpha_\mathrm{eff}$ is always found to satisfy $\alpha_\mathrm{eff} < 3$ in the fully radiative case and similarly $\alpha_\mathrm{eff} < 3/2$ in the fully adiabatic case (see Fig.~\ref{gdir_comp_full} below).

\subsection{The definition of the EQTS}

The EQTSs are surfaces of revolution about the line of sight. The general expression for their profile, in the form $\vartheta = \vartheta(r)$, corresponding to an arrival time $t_a$ of the photons at the detector, can be obtained from \citep[see e.g.][]{Brasile}:
\begin{equation}
ct_a = ct\left(r\right) - r\cos \vartheta  + r^\star\, ,
\label{ta_g}
\end{equation}
where $r^\star$ is the initial size of the expanding source, $\vartheta$ is the angle between the radial expansion velocity of a point on its surface and the line of sight, and $t \equiv t(r)$ is its equation of motion, expressed in the laboratory frame, obtained by the integration of Eqs.(\ref{Taub_Eq}). From the definition of the Lorentz gamma factor $\gamma^{-2}=1-(dr/cdt)^2$, we have:
\begin{equation}
ct=\textstyle\int_0^r\left[1-\gamma^{-2}\left(r\right)\right]^{-1/2}dr\, ,
\label{tdir}
\end{equation}
where $\gamma(r)$ comes from the integration of Eqs.(\ref{Taub_Eq}).

In the current literature, the initial size $r^\star$ is usually neglected: this is certainly possible in the description of the latest phases of the afterglow, but not in the earliest ones \citep{rbcfx02_letter}. The big difference, however, is that in our approach we use the exact solution of Eqs.(\ref{Taub_Eq}) in the evaluation of Eq.(\ref{tdir}), while in the current literature the approximate solution given in Eq.(\ref{gamma_app}) is generally used with a series of additional approximations \citep[see also][]{lett1}. All this leads to the results presented in Fig.~\ref{eqts_comp_ad} and Fig.~\ref{eqts_comp_rad}.

\subsection{The boundary conditions}

Our model, like other ones in the current literature \citep[see e.g.][]{mlr93}, is function of only two parameters: the total energy $E_{tot}$ of the electron-positron pairs originating the GRB phenomenon and their baryonic loading. In our model $E_{tot}$ coincides with the energy of the dyadosphere $E_{dya}$ \citep[see][]{rukyoto,prx98}. The expansion of the photon and electron-positron pairs pulse has been computed in full details both by semi-analytic treatments and numerical simulations \citep[the pair-electromagnetic \hbox{[PEM]} pulse, see][]{rswx99}. In our model the baryonic load is due to the engulfment by the PEM pulse of the baryonic matter left over by the collapse of the progenitor star at a radius $r\simeq 10^{10}$ cm in the still optically thick and adiabatic expansion of the pulse \citep{rswx00}. The reaching of the transparency condition of this electron-positron pair and baryonic matter is computed \citep[see e.g.][]{brx01}. The determination of the only two free parameters is then made by fitting the average intensity of the rising part, of the peak emission and of the decreasing intensity part of the afterglow \citep{lett2}. These parameters fixes uniquely the initial Lorentz gamma factor of the baryonic pulse, as well as its initial mass and the arrival time at the detector, $t_a^d\equiv(1+z)t_a$, at which the afterglow begins. As usual, $z$ is the cosmological red-shift of the source. In the specific case of GRB 991216 we have $\gamma_\circ = 310.1$, $r_\circ = 1.943 \times 10^{14}$ cm, ${(t_a^d)}_\circ = 8.413 \times 10^{-2}$ s.

\section{The treatments by Sari and Panaitescu \& M\'esz\'aros}

In the current literature \citep[see e.g.][]{pm98c} the description of the earliest parts of the afterglow are neglected and special attention is given to later times when the effect of the deceleration becomes important and Eq.(\ref{gamma_app}) is assumed to apply. The following approximations have been adopted:\\
{\bf a)} Eq.(\ref{gamma_app}) is assumed to hold only during the so-called ``deceleration phase'' when $\gamma \le \gamma_d \equiv (2/3) \gamma_\circ$ and $r > r_d$, where $\gamma=\gamma_d(r/r_d)^{-\alpha}$;\\
{\bf b)} instead of Eq.(\ref{tdir}) the following approximate expression is used:
\begin{equation}
ct=r_d\left[1+\left(4\alpha+2\right)^{-1}\gamma^{-2}\left(r\right)\right]+\textstyle\int_{r_d}^r\left[1+(1/2)\gamma^{-2}\left(r\right)\right]dr\, ;\\
\label{tdirapp}
\end{equation}
{\bf c)} no treatment of $\gamma$ is given for $r < r_d$.

\begin{figure}
\includegraphics[width=\hsize,clip]{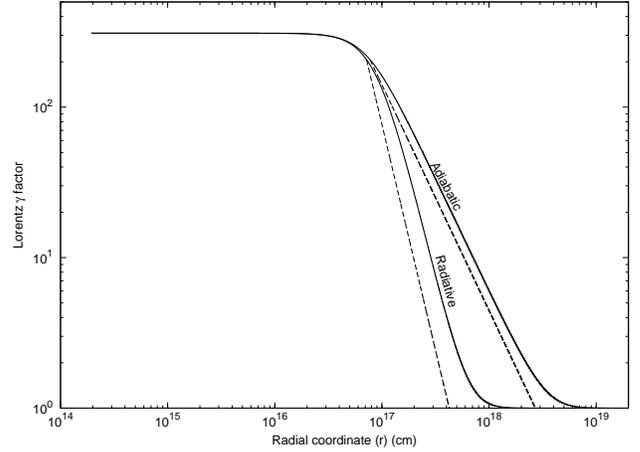}
\caption{The Lorentz gamma factors are plotted as a function of the radial coordinate both in the fully radiative and adiabatic cases for the exact (approximate) solution with solid (dashed) lines.}
\label{gdir_comp_full}
\end{figure}

\begin{figure}
\includegraphics[width=\hsize,clip]{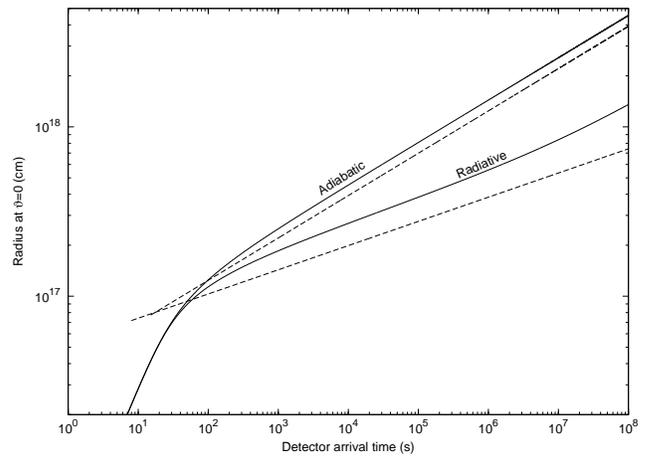}
\caption{The radius $r$ of the expanding pulse as a function of the arrival time computed at $\vartheta = 0$ is plotted both in the fully radiative and adiabatic cases for the exact (approximate) solution with solid (dashed) lines.}
\label{rdita_comp_full}
\end{figure}

In order to avoid additional arbitrary factors, we compare the results by adopting in the above approximate equations the values of $\gamma_\circ$ and $r_d$ obtained from our exact solution. In Fig.~\ref{gdir_comp_full} we compare and contrast the $\gamma$ factors as a function of the radial coordinate for the adiabatic and the fully radiative cases: the dashed (continuous) lines correspond to the approximate power-law (exact) solutions. In Fig.~\ref{rdita_comp_full} we compare and contrast the EQTS radius at $\vartheta=0$ as a function of the arrival time at the detector. Of particular interest is the crossing at $t_a^d$ smaller than $10^2$ s between the exact solution and the approximate one, which is due to the difference of about $20$ s accumulated over $10^{17}$ cm by the use of the approximate Eq.(\ref{tdirapp}) instead of the exact Eq.(\ref{tdir}). Note that this difference of $\sim 20$ s in the arrival time, in addition to the effect on the EQTSs, impedes the proper identification of the physical processes occurring in the early part of the GRB.

\begin{figure}
\includegraphics[width=\hsize,clip]{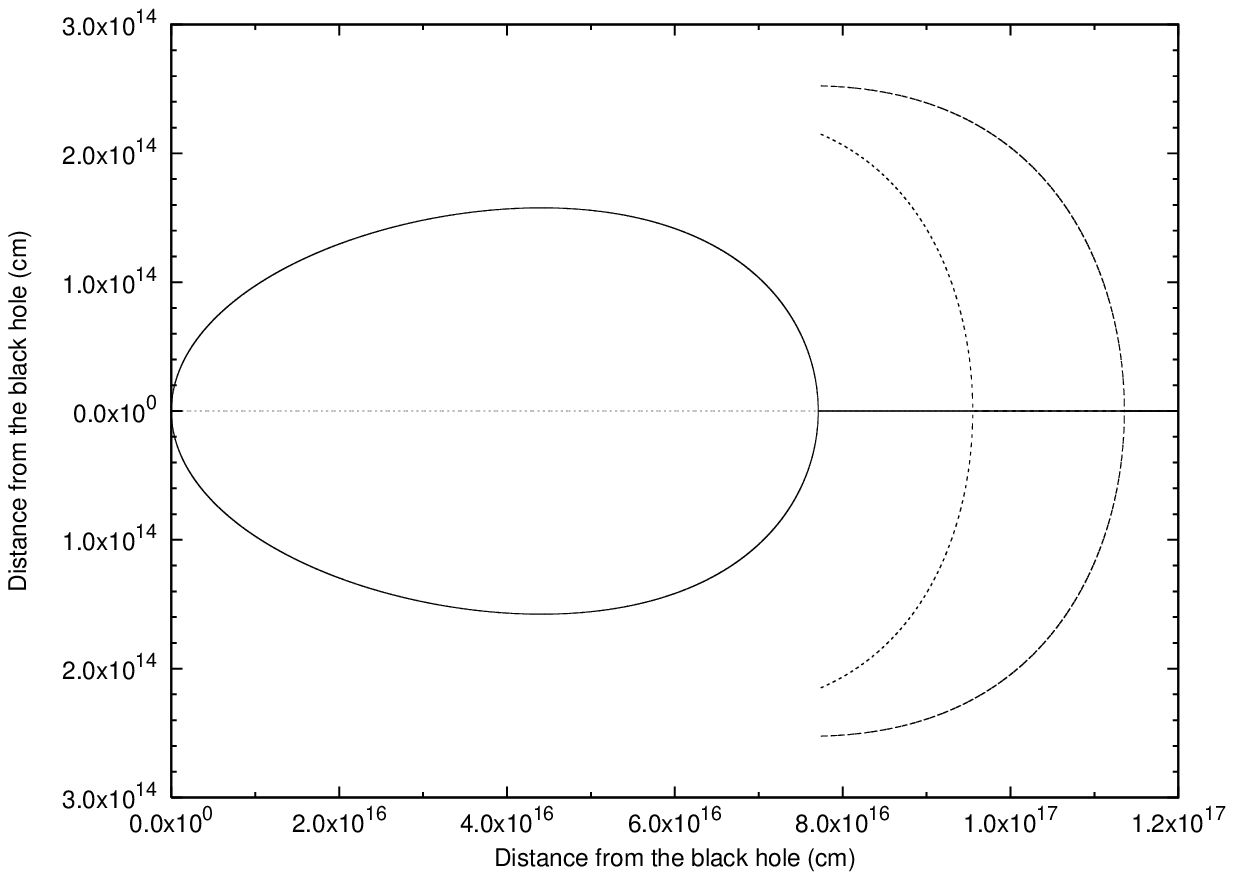}
\includegraphics[width=\hsize,clip]{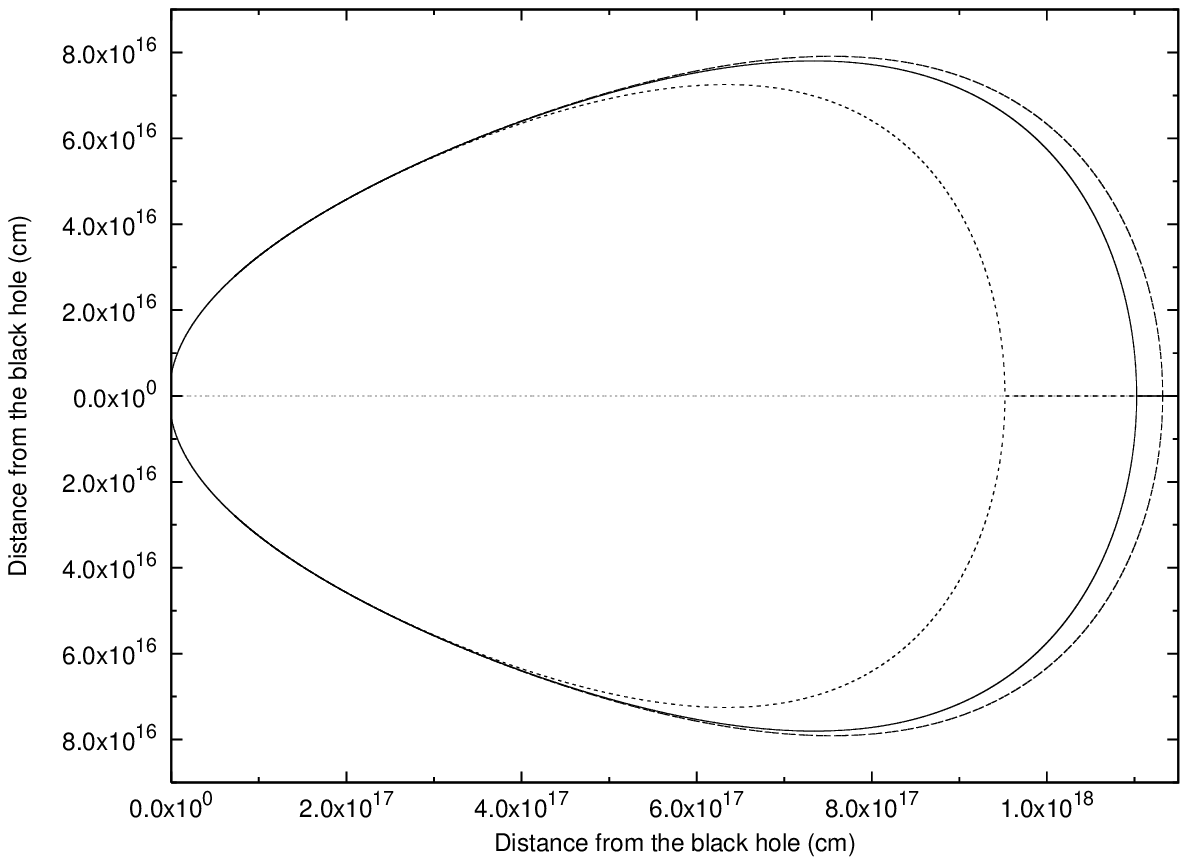}
\caption{Comparison between the EQTSs computed using the approximate formulas given by \citet{pm98c} (dotted line) and by \citet{s98,gps99} (dashed line) in the fully adiabatic case ($\alpha=3/2$ in Eq.(\ref{gamma_app}) and Eq.(\ref{tdirapp})) and the corresponding ones computed using the exact solution of the Taub Eqs.(\ref{Taub_Eq}) (solid line). The difference between the dashed line and the dotted line is due to the factor $\sqrt{2}$ in the Lorentz $\gamma$ factor adopted by Sari (see text). The upper (lower) panel corresponds to $t_a^d=35$ s ($t_a^d=4$ day).}
\label{eqts_comp_ad}
\end{figure}

\begin{figure}
\includegraphics[width=\hsize,clip]{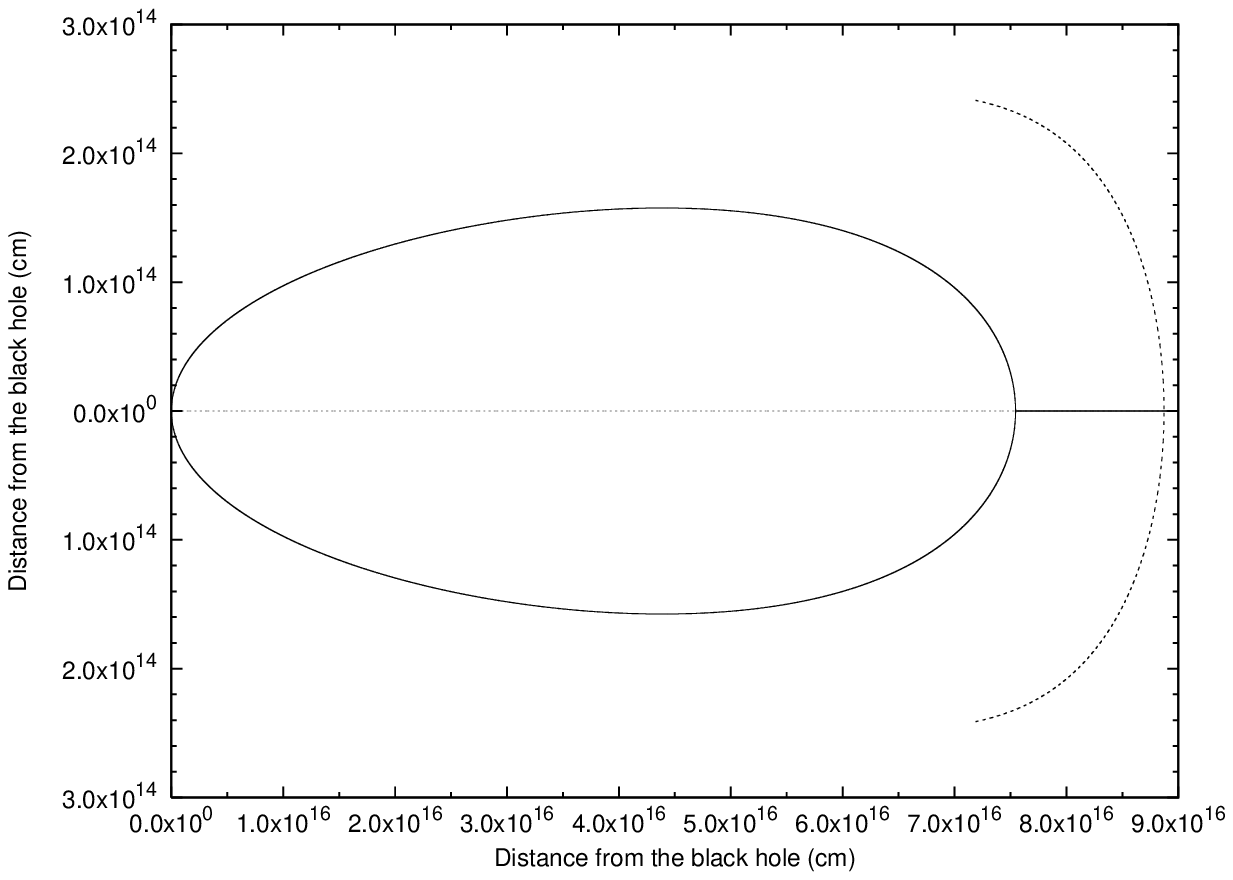}
\includegraphics[width=\hsize,clip]{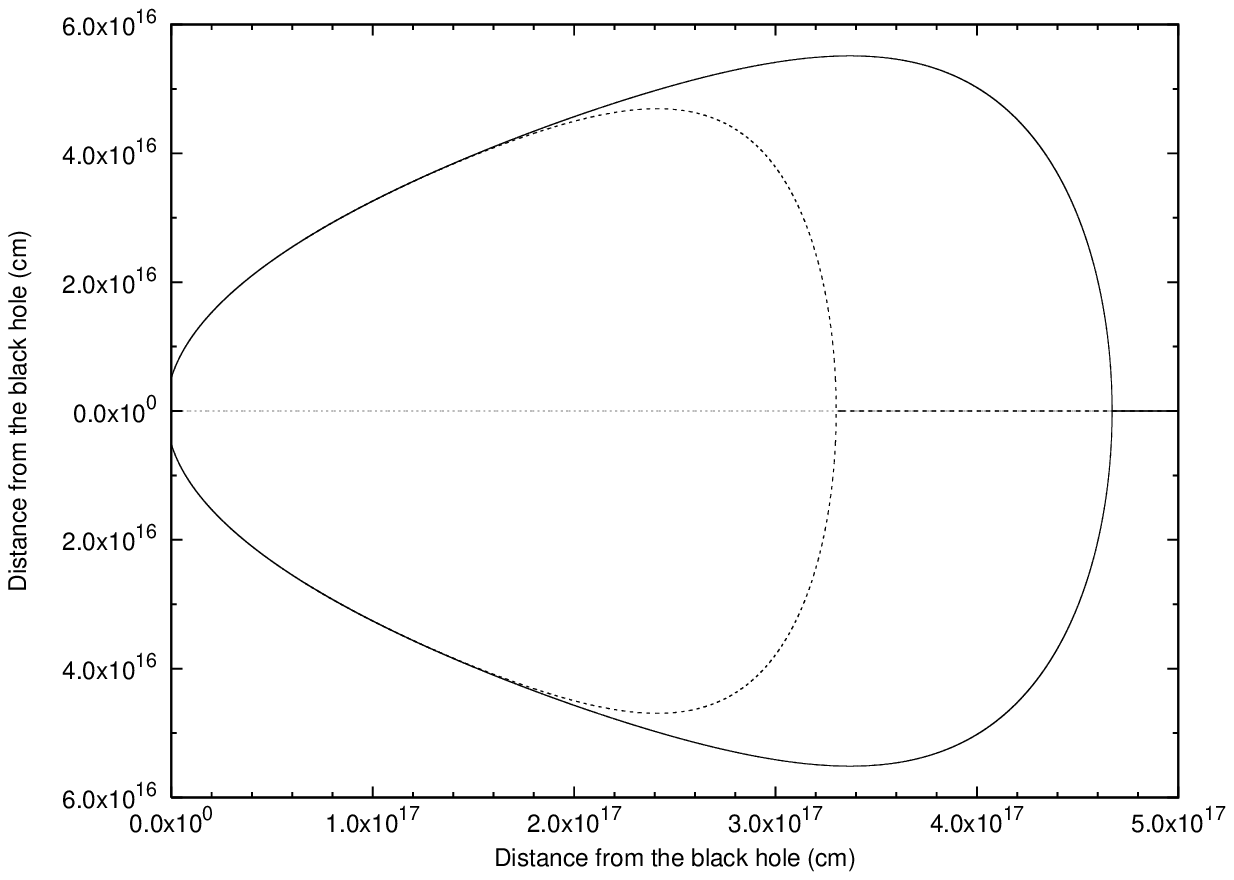}
\caption{Comparison between the EQTSs computed using the approximate formulas given by \citet{pm98c} (dotted line) in the fully radiative case ($\alpha=3$ in Eq.(\ref{gamma_app}) and Eq.(\ref{tdirapp})) and the corresponding ones computed using the exact solution of the Taub Eqs.(\ref{Taub_Eq}) (solid line). The upper (lower) panel corresponds to $t_a^d=35$ s ($t_a^d=4$ day).}
\label{eqts_comp_rad}
\end{figure}

Two different approximate treatments of the EQTSs exist in the current literature, both adopting the above approximations a), b), c): one by \citet{s98}, later used by \citet{gps99}, and another one by \citet{pm98c}. While \citet{s98} considers only the fully adiabatic case ($\alpha=3/2$ in Eq.(\ref{gamma_app}) and Eq.(\ref{tdirapp})), \citet{pm98c} consider both the fully adiabatic ($\alpha=3/2$ in Eq.(\ref{gamma_app}) and Eq.(\ref{tdirapp})) and the fully radiative ($\alpha=3$ in Eq.(\ref{gamma_app}) and Eq.(\ref{tdirapp})) cases. \citet{s98} uses a Lorentz gamma factor $\Gamma$ of a shock front propagating in the expanding pulse, with $\Gamma = \sqrt{2} \gamma$.

Without entering into the relative merit of such differing approaches, we show in Figs.~\ref{eqts_comp_ad}--\ref{eqts_comp_rad} that both of them lead to results different from the ones computed with the exact solutions.

\section{Comparison with the exact solutions and conclusions}

The consequences of using the approximate formula given in Eq.(\ref{gamma_app}) to compute the expression $t \equiv t(r)$, instead of the exact solution of the Taub Eqs.(\ref{Taub_Eq}), are clearly shown in Figs.~\ref{eqts_comp_ad}--\ref{eqts_comp_rad}. The EQTSs represented in these figures are computed at selected values of the detector arrival time both in the early ($\sim 35$ s) and in the late ($\sim 4$ day) phases of the afterglow. Both the fully radiative and fully adiabatic cases are examined. Note the approximate expression of the EQTS can only be defined for $\gamma < \gamma_d$ and $r > r_d$. Consequently, at $t_a^d=35$ s the approximate EQTSs are represented by arcs, markedly different from the exact solution (see the upper panels of Figs.~\ref{eqts_comp_ad}--\ref{eqts_comp_rad}). The same conclusion is found for the EQTS at $t_a^d=4$ days, where marked differences are found both for the fully radiative and adiabatic regimes (see the lower panels of Figs.~\ref{eqts_comp_ad}--\ref{eqts_comp_rad}).

All the observational properties of GRBs, starting from the analysis of the prompt radiation \citep{rbcfx02_letter}, to the luminosity in X- and $\gamma$-ray bands, to their spectral distribution \citep{Spectr1} as well as inferences on the possible presence or absence of beaming in GRBs, depend essentially on the structure of the EQTSs. In turn the determination of the EQTSs depends on the equations of motion of the baryonic pulse satisfying the Taub equations. The fact that the final results for the observable luminosity, spectral distribution, and substructures in the prompt radiation depend on $\sim 10^8$ integration paths on different points on the EQTSs implies that the agreement between the theoretical predictions and the observations becomes a most stringent test for the validity of the equations of motion. The correct EQTSs are also essential for the identification of the energy source of the X and $\gamma$ radiation in the GRB afterglow \citep{rbcfx02_letter,Spectr1}.

In conclusion, the approximate treatments largely overestimate (underestimate) the size of the EQTSs in the early (late) part of the afterglow. The theoretical slopes of the observables as a function of the arrival time \citep[see, e.g.,][and references therein]{p99,p00,vpkw00} are therefore incorrectly evaluated. In the meantime, analytic expressions for the EQTSs have been obtained, validating the above results and allowing the theoretical estimate of the observables in GRB afterglows \citep{br04}.

\acknowledgments

We thank the anonymous referee for constructive advices on the presentation of our results.


\begin{thebibliography}{99}

\bibitem[Bianco et al.(2001)]{brx01}
Bianco, C.L., Ruffini, R., \& Xue, S.-S. 2001, \aap, 368, 377

\bibitem[Bianco \& Ruffini(2004)]{br04}
Bianco, C.L., \& Ruffini, R. 2004, \aap, submitted to

\bibitem[Bisnovatyi-Kogan \& Murzina(1995)]{bkm95}
Bisnovatyi-Kogan, G.S., \& Murzina, M.V.A. 1995, \prd, 52, 4380

\bibitem[Blandford \& McKee(1976)]{bm76}
Blandford, R.D., \& McKee, C.F. 1976, Phys. Fluids, 19, 1130

\bibitem[Granot et al.(1999)]{gps99}
Granot, J., Piran, T., \& Sari, R. 1999, \apj, 513, 679

\bibitem[Landau \& Lif{\v s}its(1995)]{ll}
Landau, L.D., \& Lif{\v s}its, Y.M. 1995, ``Fluid Mechanics'', (Oxford: Butterworth-Heinemann), 510

\bibitem[M\'esz\'aros et al.(1993)]{mlr93}
M\'esz\'aros, P., Laguna, P., \& Rees, M.J. 1993, \apj, 415, 181

\bibitem[Panaitescu \& M\'esz\'aros(1998)]{pm98c}
Panaitescu, A., \& M\'esz\'aros, P. 1998, \apj, 493, L31

\bibitem[Piran(1999)]{p99}
Piran, T. 1999, Phys. Rep, 314, 575

\bibitem[Piran(2000)]{p00}
Piran, T. 2000, Phys. Rep., 333-334, 529

\bibitem[Preparata et al.(1998)]{prx98}
Preparata, G., Ruffini, R., \& Xue, S.-S. 1998, \aap, 338, L87

\bibitem[Ruffini(1998)]{rukyoto}
Ruffini, R. 1998, in Black Holes and High Energy Astrophysics, ed. H. Sato
\& N. Sugiyama (Tokyo: Universal Academic Press), 167

\bibitem[Ruffini et al.(2003a)]{Spectr1}
Ruffini, R., Bianco, C.L., Chardonnet, P., Fraschetti, F., Gurzadyan, V., \& Xue, S.-S. 2003a, \apjl, submitted to

\bibitem[Ruffini et al.(2003b)]{Brasile}
Ruffini, R., Bianco, C.L., Chardonnet, P., Fraschetti, F., Vitagliano, L., \& Xue, S.-S. 2003b, in AIP Conf. Proc. 668, Cosmology and Gravitation, ed. M. Novello \& S.E. Perez-Bergliaffa (Melville: AIP), 16

\bibitem[Ruffini et al.(2001a)]{lett1}
Ruffini, R., Bianco, C.L., Chardonnet, P., Fraschetti, F., \& Xue, S.-S. 2001a, \apjl, 555, L107

\bibitem[Ruffini et al.(2001b)]{lett2}
Ruffini, R., Bianco, C.L., Chardonnet, P., Fraschetti, F., \& Xue, S.-S. 2001b, \apjl, 555, L113

\bibitem[Ruffini et al.(2002)]{rbcfx02_letter}
Ruffini, R., Bianco, C.L., Chardonnet, P., Fraschetti, F., Xue, S.-S. 2002, \apjl, 581, L19

\bibitem[Ruffini et al.(1999)]{rswx99}
Ruffini, R., Salmonson, J.D.,  Wilson, J.R., \& Xue, S.S. 1999, \aap, 350, 334

\bibitem[Ruffini et al.(2000)]{rswx00}
Ruffini, R., Salmonson, J.D.,  Wilson, J.R., \& Xue, S.S. 2000, \aap, 359, 855

\bibitem[Sari(1998)]{s98}
Sari, R. 1998, \apj, 494, L49

\bibitem[Taub(1948)]{Taub}
Taub, A.H. 1948, Phys. Rev., 74, 328

\bibitem[van Paradijs et al.(2000)]{vpkw00}
van Paradijs, J., Kouveliotou, C., \& Wijers, R.A.M.J. 2000, \araa, 38, 379

\end{thebibliography}
\end{document}